\documentclass[9pt,technote]{IEEEtran}

\usepackage{cite,graphicx,psfrag,amsmath,amssymb}
\newtheorem{theorem}{Theorem}

\psfrag{mr}{$\begin{array}{l} \mbox{maximal} \\ \mbox{redundancy} \end{array}$}
\psfrag{gmr}{$\begin{array}{l} \mbox{maximal} \\ \mbox{$b$-redundancy} \end{array}$}
\psfrag{der}{$\begin{array}{l} \mbox{$d$th exponential} \\
    \mbox{redundancy} \end{array}$}
\psfrag{eHc}{$\begin{array}{r} \mbox{exponential} \\ \mbox{(Huffman)
      coding} \end{array}$}
\psfrag{pHc}{$\begin{array}{r} \mbox{bottom-merge} \\ \mbox{Huffman
      coding} \end{array}$}
\psfrag{sHc}{$\begin{array}{c} \mbox{(standard)} \\ \mbox{Huffman coding} \end{array}$}
\psfrag{-1}{$-1$}
\psfrag{0}{$0$}
\psfrag{lg54}{$\lg \frac{5}{4}$}
\psfrag{lg59}{$\lg \frac{5}{9}$}
\psfrag{lg5823}{$\lg \frac{58}{23}$}
\psfrag{pval1}{$p = (\frac{4}{9},\frac{1}{4},\frac{19}{180},\frac{1}{10},\frac{1}{10})$}
\psfrag{pval2}{$p = (0.58,0.12,0.11,0.1,0.09)$}
\psfrag{4o9}{$\frac{4}{9}$}
\psfrag{1o4}{$\frac{1}{4}$}
\psfrag{19o180}{$\frac{19}{180}$}
\psfrag{1o10}{$\frac{1}{10}$}
\psfrag{2neg1}{$2^{-1}$}
\psfrag{2negmid}{$2^{-l_{mid}}$}
\psfrag{2negL}{$2^{-L}$}
\psfrag{lmid}{$l_{mid}$}
\psfrag{n}{$n$}
\psfrag{n-m}{$n-m$}
\psfrag{L}{$L$}
\psfrag{infty}{$+\infty$}
\psfrag{1}{$1$}
\psfrag{2}{$2$}
\psfrag{3}{$3$}
\psfrag{4}{$4$}
\psfrag{8}{$8$}
\psfrag{ld}{$\ldots$}
\psfrag{li}{$l_i$}
\psfrag{ci}{$c_i$}
\psfrag{pi19}{$19 \cdot p_i$}
\psfrag{w19}{$19 \cdot w_i$}
\psfrag{wp19}{$19 \cdot w'_i$}
\psfrag{itm}{$m_i$}
\psfrag{00}{$00$}
\psfrag{01}{$01$}
\psfrag{10}{$10$}
\psfrag{11}{$11$}
\psfrag{000}{$000$}
\psfrag{001}{$001$}
\psfrag{010}{$010$}
\psfrag{011}{$011$}
\psfrag{0000}{$0000$}
\psfrag{0001}{$0001$}
\psfrag{m1}{$m_1$}
\psfrag{m2}{$m_2$}
\psfrag{m3}{$m_3$}
\psfrag{m4}{$m_4$}
\psfrag{m5}{$m_5$}
\psfrag{b}{$b$}
\psfrag{d}{$d$}

\def\boldl{{\mbox{\boldmath $l$}}}
\def\subl{{\mbox{\scriptsize \boldmath $l$}}}
\def\letterl{l}
\def\order{{O}}

\def\definedas{{}\stackrel{\Delta}{=}{}}
\def\das{{}\stackrel {{\mbox{\tiny $\Delta$}}}{{\mbox{\tiny $=$}}}{}}

\def\downto{\downarrow}
\def\upto{\uparrow}

\def\posinf{{+\infty}}

\def\lg{{\log_2}}

\def\p{{\mbox{\boldmath $p$}}}
\def\sp{{\mbox{\scriptsize \boldmath $p$}}}
\def\w{{\mbox{\boldmath $w$}}}
\def\R{{\mathbb R}}
\def\W{{\mathcal W}}
\def\X{{\mathcal X}}
\def\Z{{\mathbb Z}}
\newcommand{\defn}[0]{\it}
\begin{document}
\bibliographystyle{IEEEbib}
\title{A General Framework for Codes Involving Redundancy Minimization}
\author{Michael~B.~Baer,~\IEEEmembership{Member,~IEEE}%
\thanks{This work was supported in part by the National Science Foundation
  (NSF) under Grant CCR-9973134 and the Multidisciplinary University
  Research Initiative (MURI) under Grant DAAD-19-99-1-0215.}%
\thanks{The author was with the Department of Electrical Engineering,
  Stanford University, Stanford, CA  94305-9505 USA.  He is now with
  Electronics for Imaging, 303 Velocity Way, Foster City, CA  94404
  USA (e-mail: Michael.Baer@efi.com).}}
\markboth{IEEE Transactions on Information Theory}{A General Framework for Codes Involving Redundancy Minimization}


\maketitle

\begin{abstract}
  A framework with two scalar parameters is introduced for various
  problems of finding a prefix code minimizing a coding penalty function.
  The framework encompasses problems previously proposed by
  Huffman, Campbell, Nath, and Drmota
  and Szpankowski, shedding light on the relationships among
  these problems.  In particular, Nath's range of problems can be seen as
  bridging the minimum average redundancy problem of Huffman with the
  minimum maximum pointwise redundancy problem of Drmota and Szpankowski.
  Using this framework, two linear-time Huffman-like algorithms are devised
  for the minimum maximum pointwise redundancy problem, the only one in
  the framework not previously solved with a Huffman-like algorithm.  Both
  algorithms provide solutions common to this problem and a subrange of
  Nath's problems, the second algorithm being distinguished by its ability
  to find the minimum variance solution among all solutions common to the
  minimum maximum pointwise redundancy and Nath problems.  Simple
  redundancy bounds are also presented.
\end{abstract}


\section{Introduction} 

A source emits symbols drawn from the alphabet $\X = \{ 1, 2,
\ldots, n \}$.  Symbol $i$ has probability $p_i$, thus defining
probability mass function vector $\p$.  We assume without loss of
generality that $p_i > 0$ for every $i \in \X$, and that $p_i \leq p_j$ for
every $i>j$ ($i,j \in \X$).  The source symbols are coded into binary
codewords.  Each codeword $c_i$ corresponding to symbol $i$ has length
$\letterl_i$, thus defining length vector $\boldl$.

It is well known that Huffman coding~\cite{Huff} yields a prefix code
minimizing $\sum_{i \in \X} p_i \letterl_i$ given the natural coding
constraints: the integer constraint, $\letterl_i \in \Z_+$, and the
Kraft (McMillan) inequality~\cite{McMi}:
$$\sum_{i \in \X} 2^{-\letterl_i} \leq 1.$$  

Hu, Kleitman, and Tamaki~\cite{HKT} and Parker~\cite{Park}
independently examined other cases in which Huffman-like algorithms
were optimal; this work was later extended~\cite{Knu1,ChTh}.  Other
modifications of the Huffman coding problem were considered in
analytical papers~\cite{Camp,Nath,DrSz}, although none of these
proposed a Huffman-like algorithmic solution.  In each paper,
relationships between the modified problem and the Huffman coding
problem were explored.  Parker proposed an algorithmically-motivated
two-function parameterization defining various Huffman coding
problems; these two parameter functions are a ``weight combination''
function and a ``tree cost'' function~\cite{Park}.  Three problems,
first examined in \cite{Huff,Camp,Nath}, were considered as a part of
this framework; here we show that a fourth \cite{DrSz} fits into it as
well.  In addition, we find a simpler redundancy-motivated unifying
problem class that relates the four problems, one involving two scalar
parameters rather than two functional parameters.  This new framework
reveals a united analytical structure, including simple redundancy
bounds and novel algorithmic results which improve upon the algorithm of
\cite{DrSz}.

In Section~\ref{back}, background is given on the coding problem
introduced in \cite{Camp}.  In Section~\ref{dabr}, the new framework,
based on an extension of this problem, is introduced.  The problem and
the three other aforementioned problems are then put into the context
of this framework.  In Section~\ref{mmr}, the framework is used to
help find linear-time algorithms for the problem in \cite{DrSz}.
Redundancy bounds are presented in Section~\ref{bounds}, with
concluding thoughts following in Section~\ref{conclusion}.

\section{Background: Exponential Huffman coding}
\label{back}

One particular application of a modified coding problem was found
by Humblet~\cite{Humb2} for a problem involving minimization of buffer
overflow in communications.  In this application, the function minimized
is $\sum_{i \in \X} p_i 2^{\beta \letterl_i}$ for a given $\beta>0$.  This
is easily generalized to negative $\beta$ by specifying minimization of
the $\beta$-exponential average
\begin{equation}
F_\beta(\p,\boldl) \definedas \frac{1}{\beta} \lg \sum_{i \in \X} p_i
2^{\beta \letterl_i} \label{exphuff} .
\end{equation}
This problem was originally proposed by Campbell~\cite{Camp} and a
linear-time algorithm found independently by Hu et al. in
\cite[p.~254]{HKT}, Parker in \cite[p.~485]{Park}, and Humblet in
\cite[p.~25]{Humb0} (later published as \cite[p.~231]{Humb2}).  This
algorithm covers all of $\R$; the case of $\beta=0$ is considered by
noting that $\beta \to 0$ yields the original Huffman coding
problem.

Below is the procedure for the exponential extension of Huffman coding
with parameter $\beta$.  Note that it minimizes (\ref{exphuff})
over~$\boldl$, even if the ``probabilities'' do not add to~$1$.  We refer
to such arbitrary positive inputs as {\defn weights}, often denoted by
$\w=\{w_i\}$ instead of $\p=\{p_i\}$:

\begin{center}
{\bf Procedure for Exponential Huffman Coding}
\end{center}

\begin{enumerate}
\item Each item $m_i \in \{m_1, m_2, \ldots , m_n\}$ has weight $w_i \in
\W_{\X}$, where $\W_{\X}$ is the set of all such weights.  (Initially,
$m_i = i$.)  Assume each item $m_i$ has codeword $c_i$, to be determined
later.
\item Combine the items with the two smallest weights $w_j$ and $w_k$ into
  one item $\tilde{m}_j$ with the combined weight $\tilde{w}_j =
  2^\beta(w_j + w_k)$.  This item has codeword $\tilde{c}_j$, to be
  determined later, while $m_j$ is assigned codeword $c_j = \tilde{c}_j0$ and
  $m_k$ codeword $c_k = \tilde{c}_j1$.  Since these have been assigned in
  terms of $\tilde{c}_j$, replace $w_j$ and $w_k$ with $\tilde{w}_j$ in
  $\W$ to form $\W_{\tilde{\X}}$.
\item Repeat procedure, now with the remaining $n-1$ codewords and
  corresponding weights in $\W$, until only one item is left.  The weight
  of this item is $\sum_{i \in \X} w_i 2^{\beta l_i}$.  All codewords are
  now defined by assigning the null string to this trivial item.
\end{enumerate}

\begin{figure*}
\begin{center}
\resizebox{14cm}{!}{\includegraphics{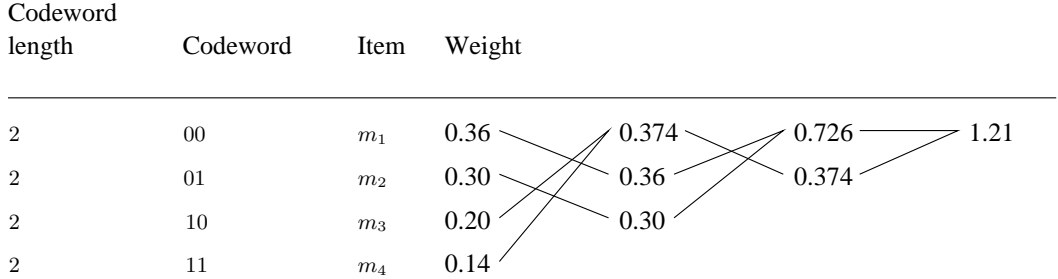}}
\caption{Exponential Huffman coding for weights $\w =
  (0.36,0.30,0.20,0.14)$ and $\beta = \lg 1.1$.  In the first step, the
  two smallest items (with weights $w_3 = 0.20$ and $w_4 = 0.14$) are
  combined into a compound item with weight $0.374 = 1.1\cdot
  (0.20+0.14)$; thus $c_3$ and $c_4$ end in $0$ and $1$, respectively.  At
  each additional step, the two smallest remaining items are combined in a
  similar fashion.  In this manner a code to optimize (and as a
  by-product calculate) the value of $\sum_{i \in \X} w_i 2^{\beta \letterl_i}$ is
  built from the bottom up.  In this case, the minimized value is $1.21$.}
\label{expfig}
\end{center}
\end{figure*}

This algorithm can be modified to run in linear time (to input size) given
sorted weights in the same manner as Huffman coding~\cite{Leeu}.  An
example of exponential Huffman coding for $\beta=\lg 1.1$ is shown in
Figure~\ref{expfig}.  The resulting code is different from that which
would be obtained via Huffman coding ($\beta=0$).

The output of a Huffman-like algorithm might be a code (and thus the
implicit code tree, e.g., \cite{Schw}) or merely codeword lengths; we
assume the latter from here on, because valid codewords can be
inferred from the lengths.  Thus we can view the problem such an
algorithm solves as an integer optimization problem.  This is useful
because many different codes can correspond to the same set of
codeword lengths and thus all be optimal for a given problem.

Considering the codeword lengths alone as the solution to a given
problem, we find that some problems have a unique optimizing set of
lengths, while others have more than one distinct optimal solution.
Multiple different solutions manifest themselves in the algorithm as
possible ties in the weight of (possibly combined) items in the
combination step (step 2 above).  Thus the algorithm, as with Huffman
coding, is nondeterministic.  Two deterministic variants are {\defn
bottom-merge Huffman coding} and {\defn top-merge Huffman
coding}~\cite{Schw}.  Code trees yielded from the former method have
been called, depending on the properties focused upon, {\defn best
Huffman trees}~\cite{Mark}, {\defn compact Huffman trees}~\cite{Hori},
{\defn minimal Huffman trees}~\cite{FoTh}, and {\defn minimum variance
Huffman trees}~\cite{Kou}, the last of these because variance is
minimized among (tied) optimal code trees (codeword lengths).

Given $b \in \R$ and $\p$, if we relax the integer constraint on $\boldl$,
minimizing $F_b(\p,\boldl)$ becomes a simple numerical optimization and
provides a lower bound for the integer-valued problem.  (We use $b$
instead of $\beta$ from here on to refer to the parameter for the
real-valued problem.)  Campbell~\cite{Camp} noted that the optimal value of
$F_b(\p,\boldl)$ for $b \in (-1,\posinf) \backslash \{0\}$ is the
R\'{e}nyi entropy of order $\alpha = (1+b)^{-1}$:
\begin{equation}
\begin{array}{rcl}
H_\alpha(\p) &\definedas& \frac{1}{1-\alpha} \lg \sum_{i \in \X} p_i^\alpha \\
&=& \frac{1+b}{b} \lg \sum_{i \in \X} p_i^{(1+b)^{-1}} .  
\end{array}
\label{Renyi} 
\end{equation}
This should not be surprising given the relationship between
Huffman coding and Shannon entropy, which corresponds to $b \to
0$, $H_1({\p})$~\cite{Shan}.

Given $b \in (-1,\posinf)$ and $\p$, the optimal ideal real-valued lengths 
achieving (\ref{Renyi}) are given by
\begin{equation}
\letterl_i^\dagger =  -\frac{1}{1+b} \lg p_i + \lg \sum_{j \in \X} {p_j}^{\frac{1}{1+b}}
. \label{idealsol1}
\end{equation}
At the extremes of the $(-1,\posinf)$ range, solutions are
defined as the limit of the solutions for $b \downto -1$ and $b \upto
\posinf$, respectively.  For $b<-1$, there is no real-valued solution, the
problem being optimized by $\letterl_1^\dagger = 0$ and
$\letterl_i^\dagger = \posinf$ for every $i>1$.

\section{Minimization of $d$-average $b$-redundancy} 
\label{dabr}

We call the difference between an integer $\letterl_i$
and the optimal real-valued solution $\letterl_i^\dagger$ 
the {\defn pointwise $b$-redundancy}
$$r_b(i) \definedas \letterl_i - \letterl_i^\dagger,$$ 
to emphasize its dependence on $b$.  The arithmetic average of
pointwise $0$-redundancy was the problem considered by Huffman in his
original paper, ``A Method for the Construction of Minimum-Redundancy
Codes.''  Here we introduce a generalization of this problem encompassing
several cases of interest.

Suppose we wish to minimize {\defn $d$-average $b$-redundancy} or {\defn
DABR}, 
\begin{equation}
R_{b,d}(\p,\boldl) \definedas \frac{1}{d} \lg \sum_{i \in \X} p_i 2^{dr_b(i)} .
\end{equation}
This amounts to finding $\boldl_{b,d}^*(\p)$ such that 
\begin{equation}
\begin{array}{rcll}
R_{b,d}(\p,\boldl_{b,d}^*(\p)) &=&
\min_\subl 
R_{b,d}(\p,\boldl) \\
&=& 
\min_\subl 
\frac{1}{d} \lg \sum_{i \in \X} p_i 2^{d(\letterl_i-\letterl_i^\dagger)} \\
 &=& \min_\subl  \frac{1}{d} \lg
\sum_{i \in \X} \frac{p_i^{\frac{1+b+d}{1+b}}}{
\sum_{j \in \X} p_j^\frac{1}{1+b}
} 2^{d\letterl_i} 
\end{array}
\label{lstar}
\end{equation}
where $\boldl$ is restricted to the integers and by the
Kraft inequality (implicit from here on).

\begin{figure}[t]
\begin{center}
\resizebox{8cm}{!}{\includegraphics{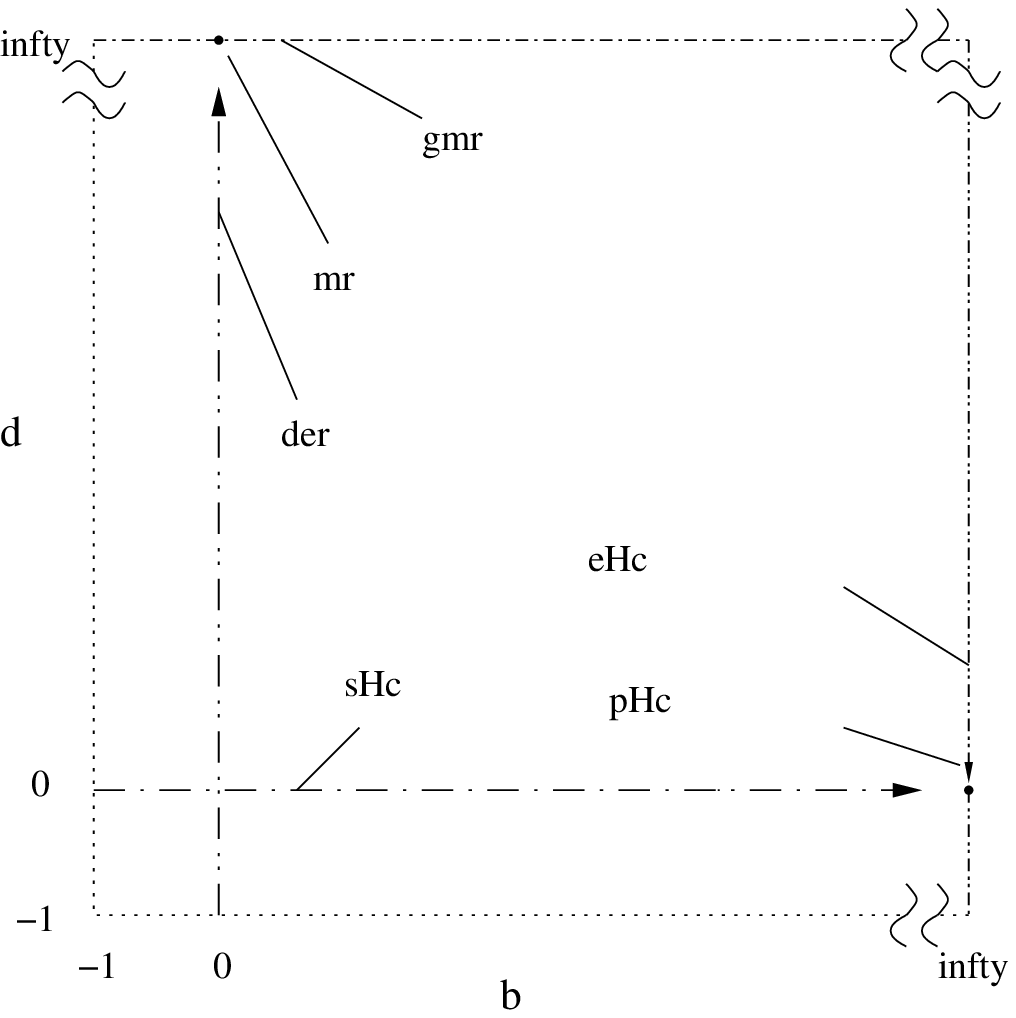}}
\caption{The parameter space for minimal $d$-average $b$-redundancy
(DABR) coding with the following noted subproblems: the Huffman
problem (the above line at $d=0$), Campbell's exponential coding
problem (line at $b=+\infty$), the problem solved via Schwartz's
bottom-merge Huffman coding method (limit point at $(+\infty,0)$ when
approached from above), Nath's $d$th exponential redundancy problem
(line at $b=0$), Drmota and Szpankowski's maximal redundancy (point at
$(0,+\infty)$), and maximal $b$-redundancy (line at $d=+\infty$).  Each
point in the extended quadrant represents a different (parameterized)
problem, as in Figure~\ref{bdfig2}.}
\label{bdfig1}
\end{center}
\end{figure}

This reduces to an exponential Huffman coding problem.  Then, given
sorted $\{p_i\}$, (\ref{lstar}) is solvable in linear time; note that
the normalization of the terms is optional for the algorithm.  For
$d<-1$, the solution is always the unary code $\boldl = (1,\ 2,\
\ldots \ ,\ n-1,\ n-1)$.  Considering the edges via limiting (as we
did with real-valued solutions), the range of nontrivial cases for
minimal DABR codes for a given probability mass function can thus be
considered to be parameterized by $b \times d \in [-1,\posinf] \times
[-1,\posinf]$, as in Figure~\ref{bdfig1}.

As indicated in this figure, many interesting coding problems fit
within this framework.  These problems, which we discuss below,
correspond to subsets of this two-dimensional extended quadrant
($[-1,\posinf] \times [-1,\posinf]$).  On the set of points for which
$b$ is $\posinf$, for example, the minimization reduces to exponential
Huffman coding with parameter $\beta=d$.  For $d=0$ ($b \in
(-1,\posinf]$) we have Huffman coding.  A particular type of Huffman
coding occurs for $b=\posinf$, $d \downto 0$.  In such a case, we note
that
\begin{equation}
R_{b,d}(\p,\boldl) = 
\sum_{i \in \X} p_i \letterl_i + \frac{d}{2}
\sigma_\sp^2({\boldl}) + \order(d^2) \qquad\hbox{as}\ d\to 0
\label{variance}
\end{equation}
where the second term on the right-hand side represents variance.  This being
the tie-breaking term, we have bottom-merge
Huffman coding.

\section{Minimization of maximal pointwise redundancy } 
\label{mmr}

As average pointwise ($0$-)redundancy has been well understood for
some time, Drmota and Szpankowski decided to explore the previously
overlooked minimization of maximal pointwise
redundancy~\cite{DrSz,DrSz2}.

We define $d$th exponential redundancy as DABR for $b=0$.  Note that
the maximal redundancy problem is equivalent to minimizing $d$th
exponential redundancy as $d\to\posinf$.  Thus, considering $d
\in [0,\posinf]$, $d$th exponential redundancy is a subproblem with a
parameter that varies solution values between minimizing average
redundancy (Huffman coding) and minimizing maximal redundancy; such a
range of problems and solutions was sought in \cite{DrSz2}.  This was
previously derived axiomatically without regard to such a range and
without solution~\cite{Nath}.  The version of the minimal DABR coding
solution applying to the maximal redundancy subproblem was found
shortly thereafter~\cite{Park}, although it was not generalized to $b
\neq 0$ or to $d=\posinf$.

Drmota and Szpankowski presented a simple method for finding a code
with minimum maximal redundancy~\cite{DrSz,DrSz2}.  However, this
solution is deficient in the following senses: First, time complexity
is $\order(n \log n)$.  Second, the Kraft inequality is not
necessarily satisfied with equality, meaning that the optimal code
found in this manner is often, in some sense, wasteful.  Third, the
code does not necessarily optimize $d$th exponential redundancy for
any $d<\posinf$.  The method is also not generalized to maximal
$b$-redundancy ($b \neq 0$).

In order to overcome the first two deficiencies, we propose a reduction to
a previously-known algorithm with linear complexity previously discussed
by Parker~\cite{Park}.  This problem was termed the {\it tree-height
measure} problem, though it was not previously considered in the context
of the maximal redundancy or DABR problems.

The tree-height measure problem minimizes the maximum value of $w_i + c
\cdot \letterl_i$ given $c > 0$ and weight vector $\w$.  Instead of using
$\tilde{w}_j = 2^\beta (w_j + w_k)$ on the merge step of Huffman coding,
the Huffman-like tree-height measure algorithm uses $\tilde{w}_j = c +
\max(w_j,w_k)$.  In order to use the tree-height measure algorithm, assign
weights according to
$$ w_i(b) = \frac{1}{1+b} \lg \frac{p_i}{p_{n}} , $$
which is always nonnegative, and let $c=1$.  Then this modified Huffman
algorithm minimizes
\begin{eqnarray*}
\max_i (w_i(b) + c \cdot \letterl_i) &=& \max_i \left(\letterl_i +
\frac{1}{1+b} \lg \frac{p_i}{p_{n}} \right) \\ &=& \max_i r_b(i) + \lg
\sum_{j \in \X} {p_j}^{\frac{1}{1+b}} \\ && - \lg p_n^{\frac{1}{1+b}} .
\end{eqnarray*}
Thus this linear-time algorithm returns a length vector
minimizing maximum $b$-redundancy and satisfying the Kraft inequality with
equality.

Because ties can occur in selecting weights to combine, the
exponential Huffman algorithm might yield one of many possible optimal
codes, including codes not optimal for the limit of $d$th exponential
redundancy (as $d \to \posinf$).  For example, consider $\p =
(\frac{8}{19},\frac{4}{19},\frac{3}{19},\frac{2}{19},\frac{2}{19})$.
For $d$th exponential redundancy, $\boldl = (1,2,3,4,4)$ and $\boldl =
(1,3,3,3,3)$ are both optimal for $d \to \posinf$.  These not
only minimize maximal redundancy, but, among codes that optimize this,
these codes also have the lowest probability of achieving this maximal
redundancy, as this is related to the second term of the expansion of
$R_{b,d}(\p,\boldl)$ for $d \to \posinf$:
\begin{equation}
\begin{array}{rcl}
R_{b,d}(\p,\boldl)
&=& \frac{1}{d} \lg \sum_{i \in \X} p_i 2^{d r_b(i)} \\
&=& \max_i r_b(i) \\
& & {+}\: \frac{1}{d} \lg P_X\left[r_b(X) = \max_j r_b(j)\right] \\
& & {+}\: \order\left(\frac{1}{d 2^d}\right) \qquad \hbox{as}\ d\to \posinf
\end{array}
\label{dinfb}
\end{equation}
Each term in the expansion has a different asymptotic complexity.  As
with minimum variance (bottom-merge) Huffman coding (\ref{variance}),
each additional term further restricts the set of feasible codes to
those that minimize the current term given the
optimization of previous terms.  In the above example, all terms are
minimized by both the aforementioned sets of lengths.  In contrast,
$\boldl = (2,2,2,3,3)$, although also minimizing maximal redundancy,
results in a code where codewords have a higher probability of achieving
maximal redundancy.  This solution, which is in some sense inferior,
can nevertheless be achieved by the tree-height measure algorithm,
specifically the bottom-merge version.

It is possible to find a $D \in \R$ such that, for every $d \geq D$,
$d$th exponential Huffman coding minimizes maximal redundancy.  Let
$\min_{i,j}^+ \gamma_{i,j}$ denote the minimum strictly positive value
of $\gamma_{i,j}$, and let $\langle x \rangle$ denote the fractional
part of $x$, i.e., $\langle x \rangle \das x - \lfloor x \rfloor$.
Assign $\delta = \min_{i,j}^+ \langle \letterl_i^\dagger -
\letterl_j^\dagger \rangle$.  It is possible to show that a
sufficiently large $D$ is given by $D = \frac{1}{\delta} \lg
\frac{2}{p_n}>1$\cite[pp.~59-62]{Baer}.  However, finding $D$ requires
sorting, so an algorithm derived from this $D$ would not be a
linear-time algorithm.

Fortunately, it is possible to arrive at a linear-time algebraic
Huffman algorithm, that is, one that keeps $D$ as a variable.
Algebraic Huffman algorithms were introduced by Knuth~\cite{Knu1}.
The one proposed here uses a Huffman algorithm which keeps track of
both the first- and second-order terms; ties between these pairs of
terms can occur only when all terms are tied, this due to the manner
in which the Huffman procedure works.  Before explaining why this is
the case, we present the algorithm.

The aforementioned first- and second-order terms are
$$w_i' \definedas \lim_{d \to \posinf} [w_i(b,d)]^{(d^{-1})}$$
and 
$$w_i'' \definedas \lim_{d \to \posinf} [w_i(b,d)]^{-1} \cdot
{[w_i']^{d}},$$ respectively, where leaf nodes have $$w_i(b,d) =
p_i^{\frac{1+b+d}{1+b}},$$ as in $d$-average $b$-redundancy.  

One can think of $w_i'$ as representing an invertible function of
maximal $b$-redundancy, $$w_i' = \left[\sum_{j=1}^n
p_j^\frac{1}{1+b}\right]^{-1} \cdot 2^{\max_i r_b(i)},$$ where, at any
given point of the algorithm, $r_b(i) = l_i - l_i^\dagger$ uses the
depth of item $i$ in its interim code tree as the value $l_i$.  Note
that only $r_b(i)$ is variable; the denominator term of $w_i'$ is a
result of not normalizing the weights at the start of the algorithm.
In a similar manner, $w_i''$ represents the probability of maximal
$b$-redundancy $P_X\left[r_b(X) = \max_j r_b(j) \right]$.

To implement this algorithm, we let $w_i'=p_i^{\frac{1}{1+b}}$ and $w_i''=p_i$
for the initial case.  In comparing items $j$ and $k$, we consider them as
lexicographically ordered pairs --- e.g., $w_j=(w_j',w_j'')$ --- so that
$w_j \geq w_k$ if and only if either $w'_j>w'_k$ or if $w'_j=w'_k$ and
$w''_j \geq w''_k$, as in \cite{Knu1}.  In combining items $j$ and $k$
(where $w_j \geq w_k$ as described), the new item will have $\tilde{w}'_j
= 2w'_j = 2 \cdot \max(w'_j,w'_k)$.  If $w'_j > w'_k$, then $\tilde{w}_j'' =
w''_j$.  Otherwise, $\tilde{w}_j'' = w''_j + w''_k$.  That is,
$$
\tilde{w}_j = 
\left\{
\begin{array}{ll}
(2w'_j, w''_j) & \mbox{ if } w'_j > w'_k \\
(2w'_j, w''_j+w''_k) & \mbox{ otherwise.} 
\end{array}
\right.
$$ The reasons for this are easily seen if we view $w_i$ as the
representation of maximal redundancy and probability this maximal
redundancy is achieved.  Take the maximum and add $1$ for the
additional bit of the codeword (multiplying $w'_i$ by 2).  Then, if
the redundancies are identical, add their probabilities ($w''_i$).
Otherwise, take the probability of the maximal redundancy.

\begin{figure*}[t]
\begin{center}
\resizebox{14cm}{!}{\includegraphics{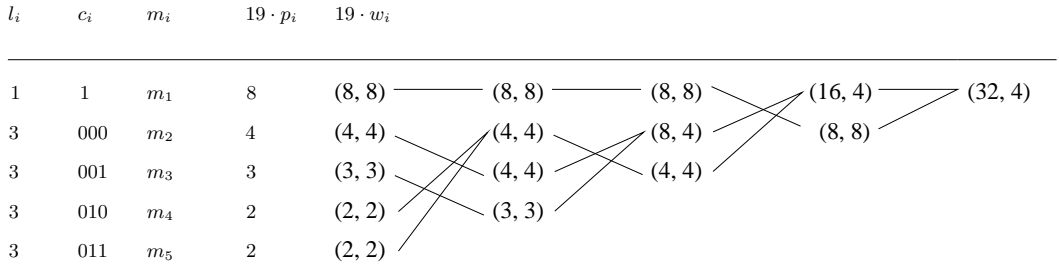}}
\caption{Algebraic maximal redundancy coding, $\p = \frac{1}{19} \cdot
  (8,4,3,2,2)$ (bottom-merge)}
\label{huffmm}
\end{center}
\end{figure*}

This combining method is a Huffman algebra, satisfying the properties
introduced in \cite{Knu1}.  The Huffman combining criterion is shown
by example in Figure~\ref{huffmm}.  The remaining weight pair after coding,
$(\frac{32}{19},\frac{4}{19})$, indicates a maximal redundancy of $\lg
\frac{32}{19}$ and a probability of $\frac{4}{19}$ that this
redundancy is achieved.

We now show that ties in the $w$ pairs imply ties in all terms of the
expansion presented in (\ref{dinfb}), or, equivalently, for $d$th
exponential redundancy for all $d \in [D, \posinf)$ where $D$ is some
(unspecified) constant.

\begin{figure*}[t]
\begin{center}
\resizebox{14cm}{!}{\includegraphics{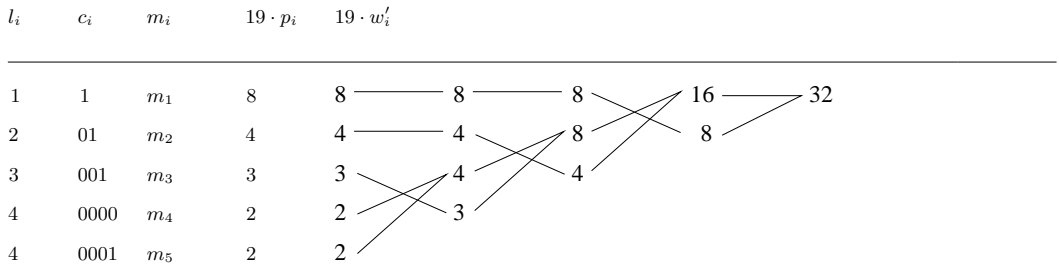}}
\caption{Top-merge maximal redundancy coding, $\p = \frac{1}{19} \cdot
  (8,4,3,2,2)$ (single variable)}
\label{hufftm}
\end{center}
\end{figure*}

\begin{theorem}
If there is a tie in the above $w$ pairs, there is a tie in all terms of
the corresponding $d$ expansion.
\end{theorem}

\begin{proof}
Consider two tied pairs.  Note that, in each,
\begin{equation}
w_i' \geq {w_i''}^{\frac{1}{1+b}}
\label{www}
\end{equation}
because this holds with equality in leaf nodes and the
inequality is preserved in the merge step, since $2 \cdot \max(a,b) \geq a+b
\geq \max(a,b)$ for $a,b \geq 0$.  If inequality (\ref{www}) holds
without equality for the tied pairs, neither node on the corresponding
code tree can be a leaf node, and, due to ordering for the combination
step, their four children must be identically weighed.  However, this
fact can be invoked inductively for either pair of children, also
tied, and thus such a tree could not be finite.  Therefore, tied pairs
arise only in cases for which the inequality holds with equality.
Thus, they must be leaf nodes or nodes with two identically-weighted
children.  Inductively, this means the subtrees must be composed of
leaf nodes that are {\defn relatively dyadic}, that is, are dyadic
when multiplied by a nontrivial common constant.  Thus they are equal
in all terms, which is what we set out to show.
\end{proof}

One can use bottom-merge or top-merge coding so that the algorithm is
deterministic.  If one uses top-merge coding --- that is, favoring
combined items over single items with identical weight~\cite{Schw} ---
one actually need not keep track of the second term; the top-merge
algorithm behaves identically without considering this term.  This
variant, illustrated in Figure~\ref{hufftm}, is actually a special
case of the tree-height measure problem mentioned above.  However, if
we wish to assure that the solution has minimum variance, the
algebraic method is needed.

\section{Bounds}
\label{bounds}

One can easily see that if we relax the integer constraint on length for
minimizing $d$-average $b$-redundancy, the real-valued solution is not
$\boldl^\dagger$, but some different $\boldl^\ddagger$.  By
substituting the solution in (\ref{idealsol1}), we find
$$ l_i^\ddagger = -\omega \cdot \lg p_i + \lg \sum_{j \in \X} p_j^{\omega} $$
where $\omega = \frac{1+b+d}{(1+b)(1+d)} =
1-\frac{bd}{(1+b)(1+d)}$.  

\begin{figure}[t]
\begin{center}
\resizebox{8cm}{!}{\includegraphics{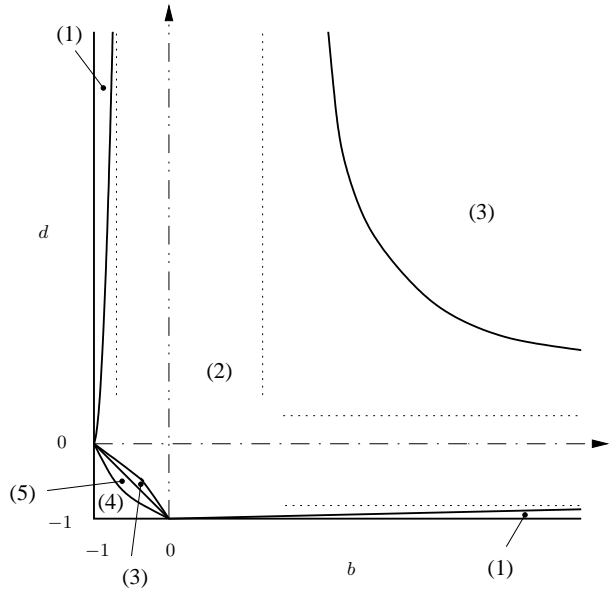}}
\caption{The parameter space for minimal DABR coding for $\p =
  (0.58,0.12,0.11,0.1,0.09)$.  Each region represents a set of problems
  with the same solution.  On the transition curves (solid), multiple
  solutions are optimal.  The five distinct solution regions are (1)
  $\boldl = (1,2,3,4,4)$, (2) $\boldl = (1,3,3,3,3)$, (3) $\boldl =
  (2,2,2,3,3)$, (4) $\boldl = (4,4,3,2,1)$, and (5) $\boldl =
  (3,3,2,2,2)$.  The dotted lines within the parameter space indicate the
  ($\posinf$) asymptotic behavior of the limits between regions.  Note
  that $b+d+1=0$ divides nonincreasing length vectors from nondecreasing.
  Also note the high degree of symmetry; the imperfection of this symmetry
  is best illustrated by the different asymptotes.}
\label{bdfig2}
\end{center}
\end{figure}

Note that when the values of $b$ and $d$ are exchanged, the ideal
solution remains the same.  This problem thus has a high degree of
symmetry.  However, because the problem itself is not symmetric, the
symmetry of integer solutions is not perfect, as we can see in
Figure~\ref{bdfig2}.  

Using this and the Shannon code~\cite{Shan} analogue $\lceil
\letterl_i^\ddagger \rceil$, we can find bounds for the optimal DABR when
$b \geq -1$, $d \geq -1$, and $b+d \geq -1$:
$$ 0 \leq R_{b,d}(\p,\boldl_{b,d}^*(\p)) - \alpha b (H_{\omega}(p) - H_{\alpha}(p)) < 1 $$
where we recall $\omega = \frac{1+b+d}{(1+b)(1+d)}$ and
$\alpha = \frac{1}{1+b}$, the subscript of R\'{e}nyi entropy in
(\ref{Renyi}).  As with exponential Huffman coding, equality holds iff
the ideal solution $\boldl^\ddagger$ has all integer lengths.  For
$b=\posinf$ and $d=0$, this results in the well-known Shannon bounds.
For $b=0$, it reduces to a normalized version of an inequality in
\cite{Nath}.  With a different normalization, this inequality
relates to R\'{e}nyi's gain of information of order $\alpha$, a
generalization of relative entropy~\cite{Ren1}.  This is not
surprising given the relationship between relative entropy and Huffman
coding noted by Longo and Galasso~\cite{LoGu}.

Due to the reduction to exponential Huffman coding, more
sophisticated redundancy results may be applied if desired.  The
bounds given by Blumer and McEliece\cite{BlMc} apply to the
exponential case but appear as solutions to related problems rather
than in closed form.  Taneja\cite{Tane} gave closed-form bounds using
an alternative definition of redundancy. 

\section{Conclusion}
\label{conclusion}

A two-dimensional framework is demonstrated to encompass examples
considered by Parker~\cite{Park} --- classical Huffman coding~\cite{Huff},
the exponential variant proposed by Campbell~\cite{Camp}, and the $d$th
exponential redundancy problem proposed by Nath~\cite{Nath}.  These
examples, along with all problems within the framework, are solvable by
Huffman-like algorithms.  The maximal redundancy problem proposed by
Drmota and Szpankowski~\cite{DrSz,DrSz2} is shown to be optimized by its
equivalence to another example considered by Parker; the top-merge version
of the algorithm in particular additionally optimizes $d$th exponential
redundancy for large $d$.  A better solution --- one minimizing codeword
length variance among such optimal codes --- is suggested by and is
developed from the two-dimensional framework introduced here.  All
algorithms discussed are Huffman-like and thus linear-time given sorted
input, unlike the original algorithm proposed for maximal redundancy.

It is unclear whether all nontrivial problems within Parker's more general
framework are covered by this seemingly more specific framework and trivial
extensions thereof.  Such analysis, building upon Parker's work, could be
a basis for further research.  Extending this algorithm to
alphabetic codes (alphabetic search trees) could also be explored.  For
nonnegative exponents ($d \geq 0$), this framework is a trivial extension of
\cite{HKT}, but negative exponents might provide more of a challenge.

\section*{Acknowledgements}

The author wishes to thank Thomas Cover, John Gill, Andrew Brzezinski,
and the two anonymous reviewers for their helpful comments and
encouragement on this paper.

\begin{biography}{Michael B. Baer} received the Bachelor's degree 
in electrical engineering and computer science from the University of
California, Berkeley, CA, in 1997 and the M.S. and the Ph.D. degrees in
electrical engineering from Stanford University, Stanford, CA, in 2000 and
2003, respectively.

From 2003 to 2005, he was a Research Scientist with UtopiaCompression, Los
Angeles, CA.  Since 2005, he has been with Electronics for Imaging, Foster
City, CA, where he is a Lead Scientist in the Imaging Group.  His research
interests include source coding trees, optimizing prefix codes for
nontraditional codeword length functionals, and image, video, and data
signal processing for compression and other applications.
\end{biography}


\begin{thebibliography}{10}

\bibitem{Huff}
D.A.~Huffman,
\newblock ``A method for the construction of minimum-redundancy codes,''
\newblock {\em Proceedings of the IRE}, vol. 40, no. 9, pp. 1098--1101, Sep.
  1952.

\bibitem{McMi}
B.~McMillan,
\newblock ``Two inequalities implied by unique decipherability,''
\newblock {\em IRE Transactions on Information Theory}, vol. IT-2, no. 4, pp.
  115--116, Dec. 1956.

\bibitem{HKT}
T.C. Hu, D.J. Kleitman, and J.K. Tamaki,
\newblock ``Binary trees optimum under various criteria,''
\newblock {\em SIAM Journal of Applied Mathematics}, vol. 37, no. 2, pp.
  246--256, Apr. 1979.

\bibitem{Park}
D.S. Parker, Jr.,
\newblock ``Conditions for Optimality of the {Huffman} algorithm,''
\newblock {\em SIAM Journal on Computing}, vol. 9, no. 3, pp. 470--489, Aug.
  1980,
\newblock Erratum: vol.~27, no.~1, Feb.~1998, p.~317.

\bibitem{Knu1}
D.E. Knuth,
\newblock ``{Huffman's} algorithm via algebra,''
\newblock {\em Journal of Combinatorial Theory, Ser.~A}, vol. 32, pp. 216--224,
  1982.

\bibitem{ChTh}
C.~Chang and J.~Thomas,
\newblock ``{Huffman} algebras for independent random variables,''
\newblock {\em Discrete Event Dynamic Systems}, vol. 4, no. 1, pp. 23--40, Feb.
  1994.

\bibitem{Camp}
L.L. Campbell,
\newblock ``Definition of entropy by means of a coding problem,''
\newblock {\em Zeitschrift f{\"{u}}r Wahrscheinlichkeitstheorie und verwandte
  Gebiete}, vol. 6, pp. 113--118, 1966.

\bibitem{Nath}
P.~Nath,
\newblock ``On a coding theorem connected with {R{\'{e}}nyi} entropy,''
\newblock {\em Information and Control}, vol. 29, no. 3, pp. 234--242, Nov.
  1975.

\bibitem{DrSz}
M.~Drmota and W.~Szpankowski,
\newblock ``Precise minimax redundancy and regret,''
\newblock {\em IEEE Transactions on Information Theory}, vol. IT-50, no. 11,
  pp. 2686--2707, Nov. 2004.

\bibitem{Humb2}
P.A. Humblet,
\newblock ``Generalization of {Huffman} coding to minimize the probability of
  buffer overflow,''
\newblock {\em IEEE Transactions on Information Theory}, vol. IT-27, no. 2, pp.
  230--232, Mar. 1981.

\bibitem{Humb0}
P.A. Humblet,
\newblock {\em Source coding for communication concentrators},
\newblock Ph.D. thesis, Massachusetts Institute of Technology, 1978.

\bibitem{Leeu}
J.~{van Leeuwen},
\newblock ``On the construction of {Huffman} trees,''
\newblock in {\em Proceedings of the 3rd International Colloquium on Automata,
  Languages, and Programming}, 1976, pp. 382--410.

\bibitem{Schw}
E.S. Schwartz,
\newblock ``An optimum encoding with minimum longest code and total number of
  digits,''
\newblock {\em Information and Control}, vol. 7, no. 1, pp. 37--44, Mar. 1964.

\bibitem{Mark}
G.~Markowsky,
\newblock ``Best {Huffman} trees,''
\newblock {\em Acta Informatica}, vol. 16, pp. 363--370, 1981.

\bibitem{Hori}
Y.~Horibe,
\newblock ``Remarks on `compact' {Huffman} trees,''
\newblock {\em Journal of Combinatorics, Information and System Sciences}, vol.
  9, no. 2, pp. 117--120, 1984.

\bibitem{FoTh}
G.~Forst and A.~Thorup,
\newblock ``Minimal {Huffman} trees,''
\newblock {\em Acta Informatica}, vol. 36, no. 9/10, pp. 721--734, Apr. 2000.

\bibitem{Kou}
L.T. Kou,
\newblock ``Minimum variance {Huffman} codes,''
\newblock {\em SIAM Journal on Computing}, vol. 9, no. 1, pp. 138--148, Feb.
  1982,
\newblock Original as {\it Minimal Variance {Huffman} Codes}, Research report
  RC 8333, International Business Machines Corporation, 1980.

\bibitem{Shan}
C.E. Shannon,
\newblock ``A mathematical theory of communication,''
\newblock {\em Bell System Technical Journal}, vol. 27, pp. 379--423, Jul.
  1948.

\bibitem{DrSz2}
M.~Drmota and W.~Szpankowski,
\newblock ``Generalized {Shannon} code minimizes the maximal redundancy,''
\newblock in {\em Proceedings of the Latin American Theoretical Informatics
  (LATIN) 2002}. 2002, pp. 306--318, Springer-Verlag.

\bibitem{Baer}
M.~B.~Baer,
\newblock {\em Coding for General Penalties},
\newblock Ph.D. thesis, Stanford University, 2003.

\bibitem{Ren1}
A.~R{\'{e}}nyi,
\newblock ``Some fundamental questions of information theory,''
\newblock {\em Magyar Tudom{\'{a}}nyos Akad{\'{e}}mia III. Osztalyanak
  K{\"{o}}zlemenei}, vol. 10, no. 1, pp. 251--282, 1960.

\bibitem{LoGu}
G.~Longo and G.~Galasso,
\newblock ``An application of informational divergence to {Huffman} codes,''
\newblock {\em IEEE Transactions on Information Theory}, vol. IT-28, no. 1, pp.
  36--43, Jan. 1982.

\bibitem{BlMc}
A.C. Blumer and R.J. McEliece,
\newblock ``The {R\'{e}nyi} redundancy of generalized {Huffman} codes,''
\newblock {\em IEEE Transactions on Information Theory}, vol. IT-34, no. 5, pp.
  1242--1249, Sep. 1988.

\bibitem{Tane}
I.J. Taneja,
\newblock ``A short note on the redundancy of degree {$\alpha$},''
\newblock {\em Information Sciences}, vol. 39, no. 2, pp. 211--216, Sep. 1986.

\end{thebibliography}
\end{document}